\documentclass[aps,prl,tightenlines,twocolumn,nofootinbib,showpacs,
letterpaper]{revtex4-1}
\usepackage{amsmath,amssymb,amsfonts}
\allowbreak
\allowdisplaybreaks
\usepackage{graphicx}
\usepackage{hyperref} 

\begin{document}
\title{Screening Masses in Gluonic Plasma }

\author{Purnendu Chakraborty$^1$, Munshi G. Mustafa$^2$ and 
Markus H. Thoma$^3$}

\affiliation{$^1$Physical Research Laboratory, Navrangpura, Ahmedabad 380009,
 India}
\affiliation{$^2$Theory Division, 
Saha Institute of Nuclear Physics, 1/AF Bidhannagar, 
Kolkata 700064, India}
\affiliation{$^3$Max-Planck-Institut f\"ur extraterrestrische Physik,
Giessenenbachstr. 85748 Garching,Germany}

\date{\today}

\pacs{11.10.Wx,12.38.Lg,12.38.Mh,25.75.Nq}

\begin{abstract}
Both electric and magnetic screening masses in a nonperturbative 
gluonic background are investigated using operator product expansion.
The magnetic screening mass is found to agree with lattice
results whereas the electric screening mass is somewhat smaller than the one found on the lattice.
\end{abstract}

\maketitle  
The long-range properties of
a thermal system are characterized by screening masses or the inverse of 
equal-time 
correlation lengths. For a given theory the screening masses determine 
the infrared sensitivity  of various thermodynamic
quantities as well as the spectral properties  of the system~\cite{fetter2003quantum}. 
In QED the screening associated with
the electric fields exhibits a non-vanishing Debye mass whereas that with  
the magnetic fields does not show up due to gauge invariance. This indicates 
that the longest length scale in hot QED plasma is dominated by magnetic 
fields. On the other hand, in QCD 
the scenario is far more complicated than in QED due to the gauge dependence
of the chromo-electric and magnetic fields, which leads to subtleties
in the calculations. 

The structure of QCD, at least near to phase transition, seems more complex than one usually expects.  
Perturbative predictions are upset by the presence of strong non-perturbative effects~\cite{Andersen:2011ug}. 
The nonperturbative determination of screening masses has been
performed in lattice QCD (LQCD )~\cite{Nakamura:2003pu,Kaczmarek:2005jy,Cucchieri:2000cy}
with an appropriate gauge fixing. The data are 
consistent with an over all exponential behaviour for the electric screening function 
in all gauges whereas that for the magnetic sector involves a nontrivial 
behaviour~\cite{Nakamura:2003pu,Cucchieri:2011di}.  
In perturbation theory the electric screening mass to the lowest order  
is obtained as $ m_D\sim gT$ (the strong coupling constant is  $\alpha_s=g^2/4\pi$ and 
$T$ is the temperature), which  falls short of the 
nonperturbative description. On the other hand, magnetic screening 
cannot be addressed in perturbation theory but one expects the magnetic
mass to be generated ($m_m\sim g m_D\sim g^2T$) 
nonperturbatively in the static sector~\cite{alex:1995nr, * Biro:1992wg,Kraemmer:2003gd}. Nonetheless, perturbative 
methods could only be reliable for temperatures far above the critical
temperature. Moreover, the perturbative power-counting hierarchy of 
scales $m_D > m _m\sim g^2 T$,  
is doubtful close to the critical temperature. Effective models using Polyakov Loop correlation~\cite{Arnold:1995bh}, 
dimensionally reduced QCD~\cite{Laine:2009dh}, ${\cal N}=4$ 
supersymmetric Yang-Mills theory~\cite{bak:2007fk} and AdS/QCD~\cite{Veschgini:2010ws}
were employed to analyze the screening masses from gauge invariant correlators. 
We note that the contributions from 
the nonperturbative magnetic sector reveal a strong dependence on these correlation functions 
but provide very useful information after all.

It is widely accepted that the nonperturbative dynamics of QCD is 
signaled by the emergence of power corrections in physical
observables. These nonperturbative corrections are 
introduced via non-vanishing vacuum expectation values  of local 
quark and gluonic operators such 
as the quark condensate \ensuremath{\langle \bar{\psi} \psi \rangle}  and 
the gluon condensate \ensuremath{\langle  \mathcal{G}_{\mu\nu}^a \mathcal{G}^{a \mu\nu} \rangle},
which are also measured in LQCD~\cite{Boyd:1996bx}.
This approach of the operator product expansion (OPE) has met noticeable 
success in QCD sum rule calculations at zero 
temperature~\cite{narison2004qcd,* shifman1992vacuum} and the calculation of the  $N$-point 
functions in QCD at zero 
temperature~\cite{Lavelle:1988eg, * Lavelle:1990xg, * Lavelle:1988fj, * 
Bagan:1989dr}. 
Unlike QCD sum rules, the condensates do not appear in a gauge invariant 
combination in QCD Green's Function~\cite{Lavelle:1992yh}. In addition, there is an
explicit dependence on the gauge fixing parameter ($\xi$) in the 
Wilson coefficients. 
The OPE has also provided some insight on the
nonperturbative features of QCD at finite temperature. 
In-medium nonperturbative
chiral quark propagator~\cite{Schaefer:1998wd}, quark-photon 
vertex~\cite{Mustafa:1999jz} and dilepton production rate~\cite{Lee:1998nz, *Mustafa:1999dt} in 
presence of dimension four electric and  magnetic condensates were investigated some 
time ago. Also attempts 
were made to extract a nonperturbative electric screening mass from the gluon 
propagator~\cite{Schmidt:1999je}. In this work we calculate for the 
first time the structure of the screening masses in a 
very comprehensive way using OPE that provides us the nonperturbative 
information to the infrared sensitivity of QCD, which in turn may help 
in constraining the various thermodynamic and spectral properties of high temperature QCD matter. 

In QCD the gluon polarization operator is not 
transverse in general,  \(P^\mu \Pi_{\mu \nu}\left(P\right) \neq 0\). 
The most general tensorial structure of the in-medium gluon self-energy 
for an $O(3)$ invariant gauge fixing condition can be 
written as~\cite{Heinz:1986kz}  
\begin{equation}
\label{self_energy_general}
\pi_{\mu\nu} \left(\omega, p\right) = \pi_l 
\mathcal{P}^l_{\mu\nu}  +  \pi_t \mathcal{P}^t_{\mu\nu} + 
\pi_m  \mathcal{M}_{\mu\nu} + 
\tilde{\pi}  \mathcal{L}_{\mu\nu}\,. 
\end{equation}
We work here in covariant gauge and also omit the color indices  for brevity.
The Lorentz invariant single particle energy and momentum are, respectively, 
given as $\omega = u \cdot P$ and 
$p =\sqrt{\left(u \cdot P\right)^2 - P^2}$ where $u^\mu$ 
is the four velocity of the heat bath and $P = \left(p_0,\vec{p}\right)$. The projection 
operators are defined 
as~\cite{Weldon:1982aq,Heinz:1986kz} 
\begin{subequations}
\begin{align}
\mathcal{P}^l_{\mu\nu} 
&= \frac{P^2}{\widetilde{P}^2}
 \bar{u}^\mu \bar{u}^\nu \,, \, \,
\mathcal{P}^t_{\mu\nu} = \eta_{\mu\nu} - u_\mu u_\nu
- \frac{\widetilde{P_\mu} \widetilde{P_\nu}}{\widetilde{P}^2}\,, \nonumber \\
\mathcal{M}_{\mu\nu} &= - \frac{1}{\sqrt{-2 \widetilde{P}^2}} \left(
\bar{u}_\mu P_\nu + \bar{u}_\nu P_\mu
\right)\,, \,
\mathcal{L}_{\mu\nu} = \frac{P_\mu P_\nu}{P^2}\, , \nonumber
\end{align} 
\end{subequations}
with  \ensuremath{\widetilde{P_\mu} = P_\mu - \omega u_\mu} and $\bar{u}^\mu
= u^\mu - ({\omega}/{P^2})P^\mu$. Both $\mathcal{P}_l^{\mu \nu}$
and $\mathcal{P}_t^{\mu \nu}$ are transverse with respect to $P^\mu$, while 
$\mathcal{M}^{\mu \nu}$ satisfies a weaker condition 
$P_\mu \mathcal{M}^{\mu \nu} P_\nu = 0$. The scalar functions 
in (\ref{self_energy_general}) are extracted as
\begin{subequations}
\begin{align}
\pi_l &= \mathcal{P}_l^{\mu\nu} \pi_{\mu\nu}\,, \, \,
\pi_t  = \frac{1}{2} \mathcal{P}_t^{\mu\nu} \pi_{\mu\nu}\,, \, \nonumber \\ 
\pi_m  &= - \mathcal{M}^{\mu\nu} \pi_{\mu\nu}\,, \, \,
\widetilde{\pi}  = \mathcal{L}^{\mu\nu} \pi_{\mu\nu}\,. \nonumber
\end{align} 
\label{pi_structures}
\end{subequations}
Here \ensuremath{\pi_m} and \ensuremath{\widetilde{\pi}} 
measure the deviation from transversality. In high temperature 
perturbative QCD (pQCD), the transversality holds only in the temporal axial gauge  and
Feynman gauge~\cite{Kraemmer:2003gd}. In general the violation of transversality is however 
sub-leading in temperature  in pQCD and 
one usually neglects 
\ensuremath{\pi_m} and \ensuremath{\tilde{\pi}}. 

Now from (\ref{self_energy_general}), the most general form of the gluon 
propagator $\mathcal{D}_{\mu\nu} = \mathcal{D}_{0,\mu\nu}\left(1 +
\pi_{\mu\nu} \mathcal{D}_{0,\mu\nu}\right)^{-1}$ follows as  
\begin{eqnarray}
&\mathcal{D}_{\mu\nu} 
= - \frac{\mathcal{P}_{\mu\nu}^t}{P^2 - \pi_t} - {2}\left[{2\left(P^2
  - \pi_l\right)\left(\xi^{-1}P^2 - \widetilde{\pi}\right) + 
\pi_m^2}\right]^{-1} \nonumber \\
& \times  \left[\left(\xi^{-1} P^2 - \widetilde{\pi}\right) 
\mathcal{P}^l_{\mu\nu} + \pi_m \mathcal{M}_{\mu\nu} + \left(P^2 - \pi_l\right)
\mathcal{L}_{\mu\nu}\right]\,.
\label{non_tr_prop}
\end{eqnarray}
The Slavnov-Taylor identity (STI) in covariant gauge, 
$P^\mu \mathcal{D}_{\mu\nu} P^\nu =  P^\mu \mathcal{D}_{0,\mu\nu} P^\nu -\xi$,    
leaves three independent components in (\ref{non_tr_prop}). 
The most general form of the nonperturbative gluon propagator is  
\begin{eqnarray}
D^{ab,{np}}_{\mu\nu}(P) &=&  \mathcal{D}^{ab,{\rm exact}}_{\mu\nu}(P)  - 
\mathcal{D}^{ab,{\rm pert}}_{\mu\nu}(P) \nonumber \\
& =& P^l_{\mu\nu}
D_l + P^t_{\mu\nu}
D_t + \mathcal{M}_{\mu\nu} D_m
\,, \label{non-pt-gluon-prop}
\end{eqnarray}
in an obvious notation.  

The chromoelectric and chromomagnetic condensates are given by the second moment of 
the nonperturbative gluon propagator~\cite{Schaefer:1998wd}
\begin{subequations}
\label{em_condensates}
\begin{align}
\left\langle \mathbf{\mathcal E}^2  \right \rangle_T &= - T F_A
\int\,\frac{d^{3}k}{\left(2\pi\right)^3}D_l\left(0, k\right) k^2 \,, \\
\left\langle \mathbf{\mathcal B}^2 \right \rangle_T
 &= 2 T F_A
\int\,\frac{d^3k}{\left(2\pi\right)^3}D_t\left(0, k\right) k^2\, ,
\end{align}
 \end{subequations} 
where $F_A=N_c^2-1$ and $N_c$ is the number of color. Note that the frequency sum is 
only restricted to the lowest 
Matsubara mode ($k_0=0$) in the spirit of the plane wave 
method~\cite{Schaefer:1998wd}.
This is equivalent to restricting oneself to the most dominant infrared 
singular sector. Also the lowest Matsubara mode excludes the explicit 
appearance of $D_m$  in the gluonic condensates. Similarly,
the ghost condensate is given as 
\begin{equation}
\label{ghost_condenset}
\left \langle \bar{\eta} \Box \eta \right \rangle_T = T F_A
\int\,\frac{d^3k}{\left(2\pi\right)^3} G\left(0, k\right) k^2\, ,
\end{equation}
where $G$ is the nonperturbative ghost propagator. 
  
\begin{figure}[t]
\centering{
\includegraphics[width=.8\columnwidth]{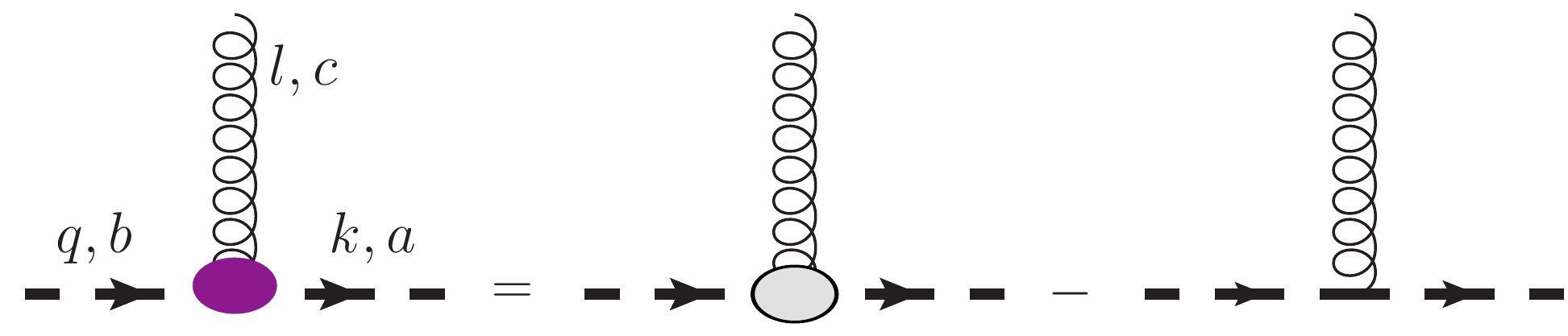}
}
\caption{\label{ghost_gluon_vertex} (color online) The nonperturbative ghost-gluon vertex}
\end{figure}

\begin{figure}[!h]
\centering{
\includegraphics[width=.25\columnwidth]{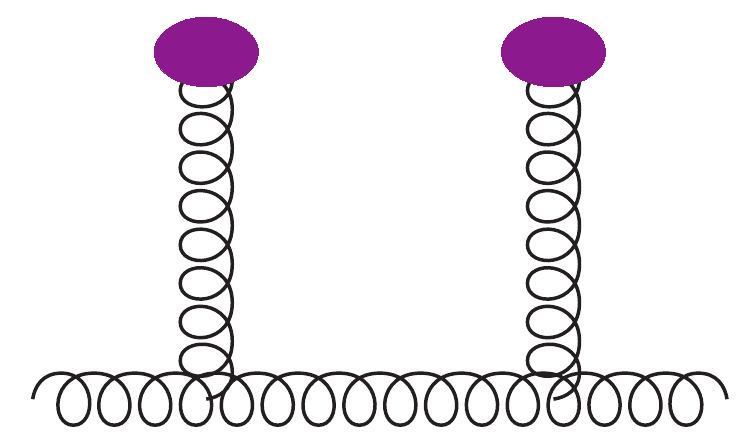}
\includegraphics[width=.25\columnwidth]{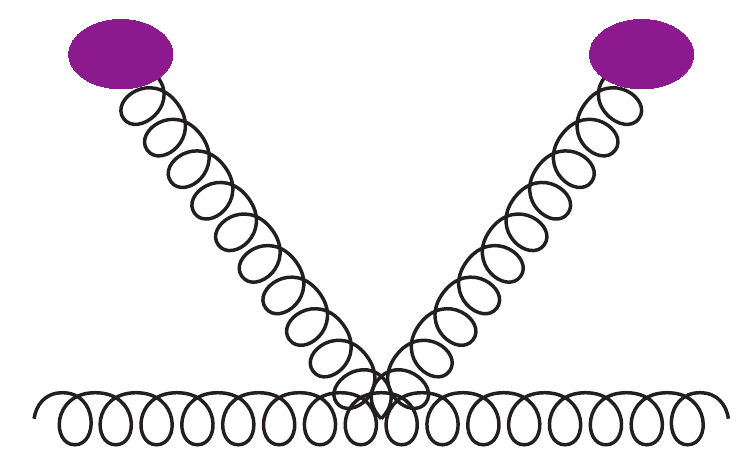}
\includegraphics[width=.25\columnwidth]{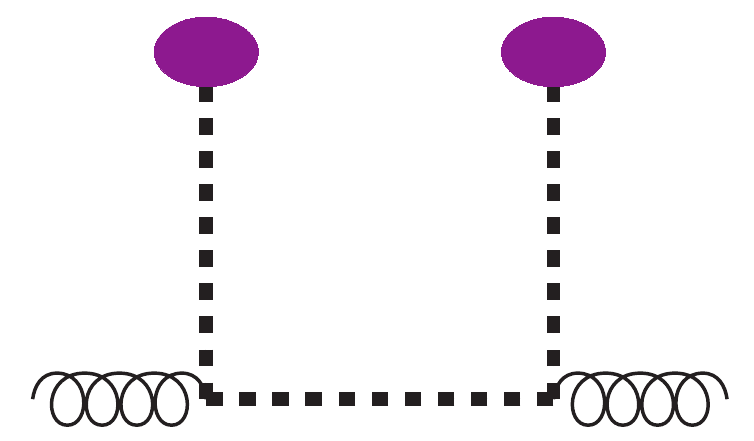}
}
\vspace*{0.2in}
\centering{
\includegraphics[width=.9\columnwidth]{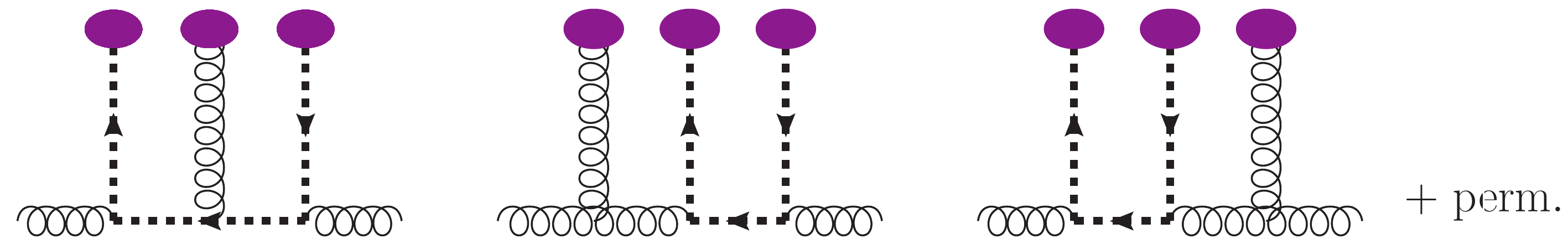}
}
\caption{\label{gluon_self1} (color online)
Gluon self-energy with gluon condensate ($1^{\rm st}$ and $2^{\rm nd}$ 
diagrams), ghost condensate ($3^{\rm rd}$ diagram) and ghost-gluon
mixed condensates ($4^{\rm th},\, 5^{\rm th}, \, 6^{\rm th}$ plus their 
permutation).}
\end{figure}

Let us note some of the important points considered in our calculations:
\begin{enumerate}
\item
We work here to the effective order of $\alpha_s$ in the sense that 
terms which are higher order in the coupling are related to terms of 
the order of  $\alpha_s$ through the equations of motion. 

\item
The gluon condensates are composite operators which do not correspond to
any conserved currents and thus, are not renormalizable. 
Nonetheless, one can extract finite non-renormalizable contributions by 
combining with other composite operators. 
The point is that under renormalization the gluonc operators 
acquire admixtures of certain other operators, {\textit e.g.,} ghost.
Therefore, we include the  
ghost-antighost condensate and two loop contributions involving 
nonperturbative ghost-gluon vertex as shown in Fig.~\ref{ghost_gluon_vertex}. 
This fixes uniquely the coefficients of dimension four 
gluonic condensates~~\cite{Spiridonov:1990rr} in the gluon self-energy
as shown in Fig.~\ref{gluon_self1}.  
The moment of the nonperturbative ghost-gluon vertex in Fig.~\ref{ghost_gluon_vertex} 
can be obtained as,
\begin{eqnarray}
 i  T^2 \int \frac{d^3k}{\left(2\pi\right)^{3}} \frac{d^3l}
{\left(2\pi\right)^{3}}
g k_i \Gamma_j^{abc} \left(k, l\right) &=& \frac{f^{abc}
  \delta_{ij}}
{3N_c F_A} \nonumber \\
\times \ \left\langle g f^{lmn}\partial_\lambda \bar{\eta}^l 
A^{\lambda,m} \eta^n \right\rangle_T  \, .
\end{eqnarray}

\item
At zero temperature, the gluon propagator that includes all possible  
condensates up to mass dimension four in Fig. \ref{gluon_self1} 
satisfies the STI where the ghost and mixed ghost-gluon 
condensates cancel the longitudinal terms generated
by the gluon condensate~\cite{Lavelle:1990xg}. However, extending it to finite temperature 
we find that the STI is not obeyed~\cite{cmt:2011b}. This is not unexpected as it is also 
found in perturbative~\cite{Kraemmer:2003gd,Heinz:1986kz} as well as in nonperturbative 
calculations~\cite{Schmidt:1999je}. 

\item
We note that the magnetic screening mass depends on the gauge fixing parameter $\xi$. On the lattice one can measure various quantities by fixing the gauge.  In the same spirit, we also intend, based on OPE with input 
from LQCD, to estimate screening masses for various gauge choices and 
compare with those lattice results.
\end{enumerate}

Now, the in-medium propagating modes can be written  
from (\ref{non_tr_prop}) as
\begin{subequations}
\begin{align}
\omega_l \left( p\right) & :\rightarrow P^2
  - \pi_l + \frac{\pi_m^2}{2\left(\xi^{-1}P^2 - \tilde{\pi}\right)}=0 \,, \\
\omega_t\left( p\right)  & :\rightarrow  P^2 - \pi_t = 0\,. 
\end{align} 
\label{tr_self_prop_mode}
\end{subequations} 
In OPE, the nonperturbative  corrections to the polarization tensor are calculated by writing down the full Feynman 
diagrams and subtracting the equivalent perturbative ones. The soft loop momenta are expanded in 
powers of external 
momenta and moments are identified with condensates as described above~\cite{Reinders:1984sr}.
This is quite different from the Hard Thermal Loop approximation of pQCD where  
the polarization operator is saturated by the hard loop momenta $(\sim T)$.  
At finite temperature, the general expressions for scalar functions in the nonperturbative gluon 
self-energy are obtained by summing all the diagrams
in Fig. \ref{gluon_self1}. These functions are quite involved~\cite{cmt:2011b}
 and reveal a rich structure of thermal QCD in a nontrivial background. 
The screening masses\footnote{In Refs.~\cite{Arnold:1995bh,Laine:2009dh} a 
different prescription was used to extract screening masses from gauge 
invariant correlators within a dimensionally reduced QCD. This dimensional 
reduction works at a rather high temperature. A direct comparison of our 
non-perturbative prescription with those in 
Refs.~\cite{Arnold:1995bh,Laine:2009dh} may not be justfied.} 
can be  extracted from the pole position of the 
propagator in the spacelike region $p_0 = 0, p^2 = - M^2$, where $M$ is the 
relevant mass scale.  The nonperturbative contribution to various scalar components of 
the gluon-self energy in the static limit $p_0\rightarrow 0$ can be obtained as
\begin{eqnarray}
\pi_l^{np} \left(0,p\right) & =& - \frac{a}{p^2}\, , \nonumber \\
\pi_t^{np} \left(0,p\right) & = & - \frac{b}{p^2}
- \frac{R \xi}{p^2}  
\left[\left\langle \bar{\eta}^a \Box \eta^a\right\rangle_T - \left\langle 
g f^{abc}  \partial_\mu \bar{\eta^a} 
A^{\mu,b} \eta^c\right\rangle_T\right]\nonumber \\
 && + \frac{R}{p^2} W_M
\left[\left\langle \bar{\eta}^a \Box \eta^a\right\rangle_T - \left\langle 
\mathbf{\mathcal B}^2\right\rangle_T + \left\langle \mathbf{\mathcal E}^2 
\right\rangle_T + 
\cdot\cdot\right], \nonumber \\
\pi_m^{np} \left(0,p\right) & =& 0, \quad \quad \tilde{\pi}^{np} \left(0,p\right) 
\neq 0\,,
\end{eqnarray}
where, 
\begin{subequations}
\begin{align}
a &= \frac{4 \pi^2  N_c}{F_A}
\left[\frac{8}{3} \frac{\alpha_s}{\pi} \left \langle 
\mathbf{\mathcal E}^2 \right \rangle_T   + 
\frac{8}{30} \frac{\alpha_s}{\pi} \left \langle 
\mathbf{\mathcal B}^2 \right \rangle_T 
\right] \,, \\
b &= \frac{4 \pi^2  N_c}{F_A}
\left[ W_E \frac{\alpha_s}{\pi} \left 
\langle \mathbf{\mathcal E}^2 \right
\rangle_T  - W_B \frac{\alpha_s}{\pi} \left 
\langle \mathbf{\mathcal B}^2 \right
\rangle_T 
\right]\,. 
\end{align}
\end{subequations}
Here  $R = \frac{4 \pi \alpha_s N_c}{3F_A}$, $W_E = \left(2 + \frac{\xi}{3}\right)$, $W_B = \frac{1}{15}\left(38 + 9\xi\right)$ 
and $W_M = \left(2+\xi\right)$. The value of $a$ obtained here is the same as in 
Ref.~\cite{Schmidt:1999je}.
Condensates appearing in the last two terms in $\pi_t$   
are classical equations 
of motion for ghost and gluon fields so they vanish.  
The electric and magnetic screening masses are obtained from  
\begin{eqnarray}
\label{eqn_mass_debye}
m_D^2 &=& \pi_l^{pert}\left(0, -m_D^2\right) + \pi_l^{np}\left(0,  -m_D^2\right)\,, \\
\label{eqn_mass_mag}
m_m^2 &=& \pi_t^{pert}\left(0, -m_m^2\right) + \pi_t^{np}\left(0, -m_m^2\right).
\end{eqnarray} 
To the perturbative order $\alpha_s$, $\pi_l^{pert}\left(0,p\right) = (m_D^{pert})^2 = 4\pi\alpha_sT^2$, whereas 
$\pi^{pert}_t\left(0,p\right) = 0$. Solving (\ref{eqn_mass_debye}) and (\ref{eqn_mass_mag}), 
we obtain the values of the screening masses as 
\begin{equation}
m_m = b^{\frac{1}{4}}\,, \quad m_D = \left[\frac{1}{2}\left\{(m_D^{pert})^2 + 
\sqrt{(m_D^{pert})^4 + 4a}\right\}\right]^{\frac{1}{2}}\,. 
\end{equation} 
For numerical evaluations of nonperturbative part in screening masses we use electric
and magnetic condensates  related to space ($\Delta_\sigma$) and 
timelike ($\Delta_\tau$) plaquettes measured on lattice for pure $SU(3)$  
gauge theory~\cite{Boyd:1996bx} as
\begin{eqnarray}
\frac{\alpha_s}{\pi}\left\langle \mathbf{\mathcal E}^2 \right \rangle_T &=& 
\frac{4}{11} T^4 \Delta_\tau - \frac{2}{11}\left\langle  
\mathbf{\mathcal G}^2 \right \rangle_0\,, \nonumber \\
\frac{\alpha_s}{\pi}\left\langle \mathbf{\mathcal B}^2 \right \rangle_T &=& 
-\frac{4}{11} T^4 \Delta_\sigma + \frac{2}{11}\left\langle  
\mathbf{\mathcal G}^2 \right \rangle_0\,,
\label{conds}
\end{eqnarray}   
where $\left\langle  \mathbf{\mathcal G}^2 \right \rangle_0$ is gluon 
condensate at $T=0$ and we take $\left\langle  \mathbf{\mathcal G}^2 \right 
\rangle_0/T_c^4  = 2.5$ and  the critical temperature, $T_c = 260$ MeV, for pure $SU(3)$ 
gauge theory. 
The perturbative piece is evaluated using the two loop running coupling constant, 
\begin{equation}
\alpha_s \left(\bar{\mu}\right) = \frac{4\pi}{\beta_0 \bar{L}}\left[1 - 
\frac{2\beta_1}{\beta_0^2} \frac{\ln{\bar{L}}}{\bar{L}}\right]\,,
\end{equation}
with $\beta_0 = 11$, $\beta_1 = 51$, $\bar{L} = \ln{\left(\bar{\mu}^2/\Lambda^2\right)}$. We take
$\bar{\mu} = 2 \pi T$ and $\Lambda = 1.03 T_c$.

We further note that there is no $\alpha_s$ dependence of the condensates in
(\ref{conds}), i.e., they cannot be expanded as a power series with respect to $\alpha_s$,  since 
they are based on non-perturtbative LQCD results~\cite{Schaefer:1998wd}.
Due to the same reason, the parametric power counting hierarchy of 
scales $m_D>m_m\sim g^2T$ is obscure at low temperature where the condensates
might provide plausible explanation for the magnetic mass. 

\begin{figure}[!h]
\centering{
\includegraphics[width=.8\columnwidth]{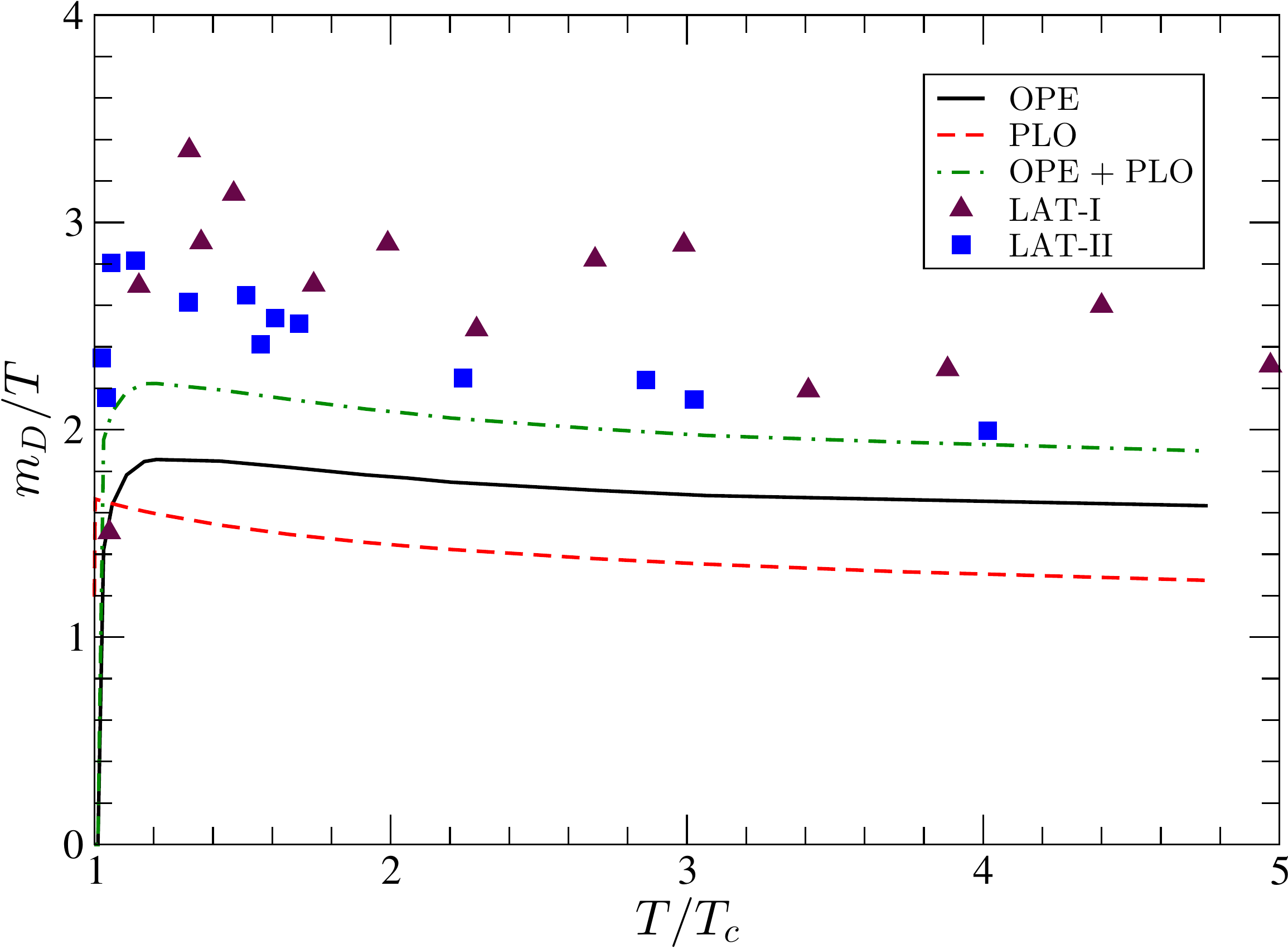}
}
\centering{
\includegraphics[width=.8\columnwidth]{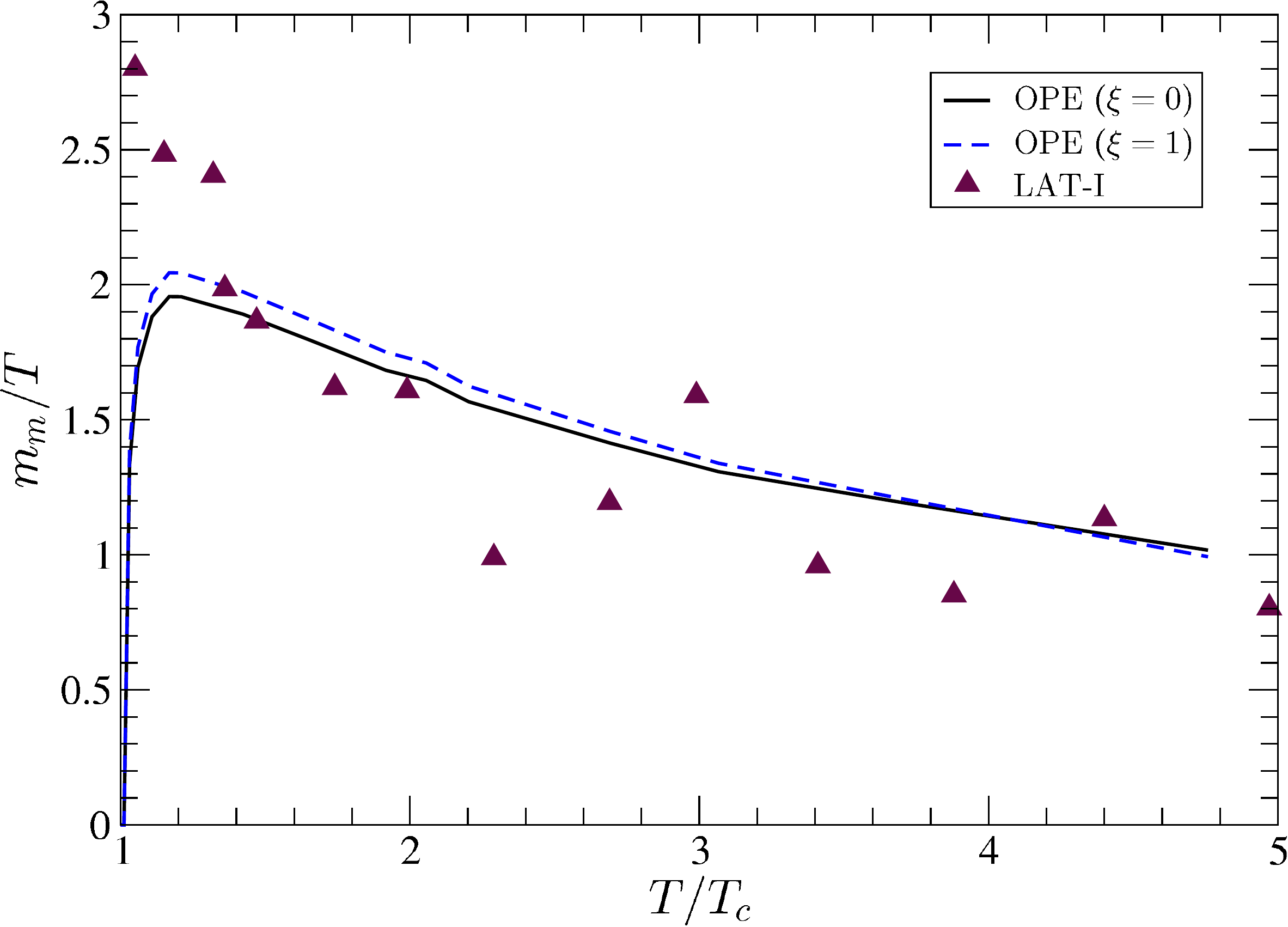}
}
\caption{\label{screen_mass} (color online) 
Temperature variation of electric (upper panel) and magnetic (lower panel)
screening masses. The present investigation is represented by OPE with 
two different gauge fixing parameter $\xi$  whereas
LQCD data are represented by LAT-I~\cite{Nakamura:2003pu} and 
LAT-II~\cite{Kaczmarek:2005jy}. PLO is perturbative leading order.
}
\end{figure}

The electric and magnetic screening masses so obtained are delineated in Fig.~\ref{screen_mass}. 
We find that the nonperturbative contribution in $m_D$ dominates over the perturbative leading order 
(PLO) contribution for the temperature range we considered. The complete 
electric screening mass, including perturbative and nonperturbative contributions,
 falls short of lattice data but is still rather close to it. On the other hand, the magnetic screening is purely 
nonperturbative in nature and agrees relatively well with lattice data. As seen the magnetic mass is dependent
 upon the gauge fixing parameter and we have chosen Landau $(\xi = 0)$  and Feynman $(\xi = 1)$ gauge. 
The weak gauge dependence found here is in agreement with that of Ref.~\cite{Nakamura:2003pu} for a similar choice of gauge fixing.       
 
In summary, we have for the first time computed the nonperturbative contribution  to both 
chromoelectric and chromomagnetic screening masses using OPE in a gluonic plasma. In particular, the magnetic screening mass is
in relatively good agreement with the LQCD data whereas that of electric screening exhibits some discrepancy. The OPE electric 
screening mass is about 20\% below the LQCD data points, which, however, show a rather large spread.
The knowledge of these quantities sets the dynamical length scale  and provides us with the 
active degrees of freedom in a hot QCD plasma. These results may be useful  input to calculate various 
thermodynamic quantities, spectral properties and for the phenomenology of 
jet quenching~\cite{Wang:2000uj}, 
quarkonium suppression~\cite{Matsui:1986dk} in hot QCD matter produced in relativistic heavy-ion collision experiments.


\begin{acknowledgements}
PC acknowledges the hospitality at CAPSS, Bose Institute and Theory Division, SINP, Kolkata, 
India where parts of the work were done.
\end{acknowledgements}
\bibliography{grf_new}{}

\begin{thebibliography}{33}%
\makeatletter
\providecommand \@ifxundefined [1]{%
 \@ifx{#1\undefined}
}%
\providecommand \@ifnum [1]{%
 \ifnum #1\expandafter \@firstoftwo
 \else \expandafter \@secondoftwo
 \fi
}%
\providecommand \@ifx [1]{%
 \ifx #1\expandafter \@firstoftwo
 \else \expandafter \@secondoftwo
 \fi
}%
\providecommand \natexlab [1]{#1}%
\providecommand \enquote  [1]{``#1''}%
\providecommand \bibnamefont  [1]{#1}%
\providecommand \bibfnamefont [1]{#1}%
\providecommand \citenamefont [1]{#1}%
\providecommand \href@noop [0]{\@secondoftwo}%
\providecommand \href [0]{\begingroup \@sanitize@url \@href}%
\providecommand \@href[1]{\@@startlink{#1}\@@href}%
\providecommand \@@href[1]{\endgroup#1\@@endlink}%
\providecommand \@sanitize@url [0]{\catcode `\\12\catcode `\$12\catcode
  `\&12\catcode `\#12\catcode `\^12\catcode `\_12\catcode `\%12\relax}%
\providecommand \@@startlink[1]{}%
\providecommand \@@endlink[0]{}%
\providecommand \url  [0]{\begingroup\@sanitize@url \@url }%
\providecommand \@url [1]{\endgroup\@href {#1}{\urlprefix }}%
\providecommand \urlprefix  [0]{URL }%
\providecommand \Eprint [0]{\href }%
\@ifxundefined \urlstyle {%
  \providecommand \doi  [0]{\begingroup \@sanitize@url \@doi}%
  \providecommand \@doi [1]{\endgroup \@@startlink {\doibase
  #1}doi:\discretionary {}{}{}#1\@@endlink }%
}{%
  \providecommand \doi  [0]{doi:\discretionary{}{}{}\begingroup
  \urlstyle{rm}\Url }%
}%
\providecommand \doibase [0]{http://dx.doi.org/}%
\providecommand \Doi [0]{\begingroup \@sanitize@url \@Doi }%
\providecommand \@Doi  [1]{\endgroup\@@startlink{\doibase#1}\@@Doi}%
\providecommand \@@Doi [1]{#1\@@endlink}%
\providecommand \selectlanguage [0]{\@gobble}%
\providecommand \bibinfo  [0]{\@secondoftwo}%
\providecommand \bibfield  [0]{\@secondoftwo}%
\providecommand \translation [1]{[#1]}%
\providecommand \BibitemOpen [0]{}%
\providecommand \bibitemStop [0]{}%
\providecommand \bibitemNoStop [0]{.\EOS\space}%
\providecommand \EOS [0]{\spacefactor3000\relax}%
\providecommand \BibitemShut  [1]{\csname bibitem#1\endcsname}%
\bibitem [{\citenamefont {Fetter}\ and\ \citenamefont
  {Walecka}(2003)}]{fetter2003quantum}%
  \BibitemOpen
  \bibfield  {author} {\bibinfo {author} {\bibfnamefont {A.}~\bibnamefont
  {Fetter}}\ and\ \bibinfo {author} {\bibfnamefont {J.}~\bibnamefont
  {Walecka}},\ }\href@noop {} {\emph {\bibinfo {title} {Quantum theory of
  many-particle systems}}}\ (\bibinfo  {publisher} {Dover Pubns},\ \bibinfo
  {year} {2003})\BibitemShut {NoStop}%
\bibitem [{\citenamefont {Andersen}\ \emph {et~al.}(2011)\citenamefont
  {Andersen}, \citenamefont {Leganger}, \citenamefont {Strickland},\ and\
  \citenamefont {Su}}]{Andersen:2011ug}%
  \BibitemOpen
  \bibfield  {author} {\bibinfo {author} {\bibfnamefont {J.~O.}\ \bibnamefont
  {Andersen}}, \bibinfo {author} {\bibfnamefont {L.~E.}\ \bibnamefont
  {Leganger}}, \bibinfo {author} {\bibfnamefont {M.}~\bibnamefont
  {Strickland}}, \ and\ \bibinfo {author} {\bibfnamefont {N.}~\bibnamefont
  {Su}},\ }\Doi {10.1103/PhysRevD.84.087703, 10.1103/PhysRevD.84.089906,
  10.1103/PhysRevD.84.087703, 10.1103/PhysRevD.84.089906} {\bibfield  {journal}
  {\bibinfo  {journal} {Phys.Rev.},\ }\textbf {\bibinfo {volume} {D84}},\
  \bibinfo {pages} {087703} (\bibinfo {year} {2011})}\BibitemShut {NoStop}%
\bibitem [{\citenamefont {Nakamura}\ \emph {et~al.}(2004)\citenamefont
  {Nakamura}, \citenamefont {Saito},\ and\ \citenamefont
  {Sakai}}]{Nakamura:2003pu}%
  \BibitemOpen
  \bibfield  {author} {\bibinfo {author} {\bibfnamefont {A.}~\bibnamefont
  {Nakamura}}, \bibinfo {author} {\bibfnamefont {T.}~\bibnamefont {Saito}}, \
  and\ \bibinfo {author} {\bibfnamefont {S.}~\bibnamefont {Sakai}},\ }\Doi
  {10.1103/PhysRevD.69.014506} {\bibfield  {journal} {\bibinfo  {journal}
  {Phys. Rev.},\ }\textbf {\bibinfo {volume} {D69}},\ \bibinfo {pages} {014506}
  (\bibinfo {year} {2004})}\BibitemShut {NoStop}%
\bibitem [{\citenamefont {Kaczmarek}\ and\ \citenamefont
  {Zantow}()}]{Kaczmarek:2005jy}%
  \BibitemOpen
  \bibfield  {author} {\bibinfo {author} {\bibfnamefont {O.}~\bibnamefont
  {Kaczmarek}}\ and\ \bibinfo {author} {\bibfnamefont {F.}~\bibnamefont
  {Zantow}},\ }\href@noop {} {}\Eprint {http://arxiv.org/abs/hep-lat/0512031}
  {arXiv:hep-lat/0512031 [hep-lat]} \BibitemShut {NoStop}%
\bibitem [{\citenamefont {Cucchieri}\ \emph {et~al.}(2001)\citenamefont
  {Cucchieri}, \citenamefont {Karsch},\ and\ \citenamefont
  {Petreczky}}]{Cucchieri:2000cy}%
  \BibitemOpen
  \bibfield  {author} {\bibinfo {author} {\bibfnamefont {A.}~\bibnamefont
  {Cucchieri}}, \bibinfo {author} {\bibfnamefont {F.}~\bibnamefont {Karsch}}, \
  and\ \bibinfo {author} {\bibfnamefont {P.}~\bibnamefont {Petreczky}},\ }\Doi
  {10.1016/S0370-2693(00)01331-9} {\bibfield  {journal} {\bibinfo  {journal}
  {Phys. Lett.},\ }\textbf {\bibinfo {volume} {B497}},\ \bibinfo {pages} {80}
  (\bibinfo {year} {2001})}\BibitemShut {NoStop}%
\bibitem [{\citenamefont {Cucchieri}\ and\ \citenamefont
  {Mendes}()}]{Cucchieri:2011di}%
  \BibitemOpen
  \bibfield  {author} {\bibinfo {author} {\bibfnamefont {A.}~\bibnamefont
  {Cucchieri}}\ and\ \bibinfo {author} {\bibfnamefont {T.}~\bibnamefont
  {Mendes}},\ }\href@noop {} {}\Eprint {http://arxiv.org/abs/1105.0176}
  {arXiv:1105.0176 [hep-lat]} \BibitemShut {NoStop}%
\bibitem [{\citenamefont {Alexanian}\ and\ \citenamefont
  {Nair}(1995)}]{alex:1995nr}%
  \BibitemOpen
  \bibfield  {author} {\bibinfo {author} {\bibfnamefont {G.}~\bibnamefont
  {Alexanian}}\ and\ \bibinfo {author} {\bibfnamefont {V.}~\bibnamefont
  {Nair}},\ }\Doi {10.1016/0370-2693(95)00475-Z} {\bibfield  {journal}
  {\bibinfo  {journal} {Phys. Lett.},\ }\textbf {\bibinfo {volume} {B352}},\
  \bibinfo {pages} {435} (\bibinfo {year} {1995})}\BibitemShut {NoStop}%
\bibitem [{\citenamefont {Biro}\ and\ \citenamefont
  {Muller}(1993)}]{Biro:1992wg}%
  \BibitemOpen
  \bibfield  {author} {\bibinfo {author} {\bibfnamefont {T.}~\bibnamefont
  {Biro}}\ and\ \bibinfo {author} {\bibfnamefont {B.}~\bibnamefont {Muller}},\
  }\Doi {10.1016/0375-9474(93)90061-2} {\bibfield  {journal} {\bibinfo
  {journal} {Nucl. Phys.},\ }\textbf {\bibinfo {volume} {A561}},\ \bibinfo
  {pages} {477} (\bibinfo {year} {1993})}\BibitemShut {NoStop}%
\bibitem [{\citenamefont {Kraemmer}\ and\ \citenamefont
  {Rebhan}(2004)}]{Kraemmer:2003gd}%
  \BibitemOpen
  \bibfield  {author} {\bibinfo {author} {\bibfnamefont {U.}~\bibnamefont
  {Kraemmer}}\ and\ \bibinfo {author} {\bibfnamefont {A.}~\bibnamefont
  {Rebhan}},\ }\Doi {10.1088/0034-4885/67/3/R05} {\bibfield  {journal}
  {\bibinfo  {journal} {Rept. Prog. Phys.},\ }\textbf {\bibinfo {volume}
  {67}},\ \bibinfo {pages} {351} (\bibinfo {year} {2004})}\BibitemShut
  {NoStop}%
\bibitem [{\citenamefont {Arnold}\ and\ \citenamefont
  {Yaffe}(1995)}]{Arnold:1995bh}%
  \BibitemOpen
  \bibfield  {author} {\bibinfo {author} {\bibfnamefont {P.~B.}\ \bibnamefont
  {Arnold}}\ and\ \bibinfo {author} {\bibfnamefont {L.~G.}\ \bibnamefont
  {Yaffe}},\ }\Doi {10.1103/PhysRevD.52.7208} {\bibfield  {journal} {\bibinfo
  {journal} {Phys. Rev.},\ }\textbf {\bibinfo {volume} {D52}},\ \bibinfo
  {pages} {7208} (\bibinfo {year} {1995})}\BibitemShut {NoStop}%
\bibitem [{\citenamefont {Laine}\ and\ \citenamefont
  {Vepsalainen}(2009)}]{Laine:2009dh}%
  \BibitemOpen
  \bibfield  {author} {\bibinfo {author} {\bibfnamefont {M.}~\bibnamefont
  {Laine}}\ and\ \bibinfo {author} {\bibfnamefont {M.}~\bibnamefont
  {Vepsalainen}},\ }\Doi {10.1088/1126-6708/2009/09/023} {\bibfield  {journal}
  {\bibinfo  {journal} {JHEP},\ }\textbf {\bibinfo {volume} {0909}},\ \bibinfo
  {pages} {023} (\bibinfo {year} {2009})}\BibitemShut {NoStop}%
\bibitem [{\citenamefont {Bak}\ \emph {et~al.}(2007)\citenamefont {Bak},
  \citenamefont {Karch},\ and\ \citenamefont {Yaffe}}]{bak:2007fk}%
  \BibitemOpen
  \bibfield  {author} {\bibinfo {author} {\bibfnamefont {D.}~\bibnamefont
  {Bak}}, \bibinfo {author} {\bibfnamefont {A.}~\bibnamefont {Karch}}, \ and\
  \bibinfo {author} {\bibfnamefont {L.~G.}\ \bibnamefont {Yaffe}},\ }\Doi
  {10.1088/1126-6708/2007/08/049} {\bibfield  {journal} {\bibinfo  {journal}
  {JHEP},\ }\textbf {\bibinfo {volume} {0708}},\ \bibinfo {pages} {049}
  (\bibinfo {year} {2007})}\BibitemShut {NoStop}%
\bibitem [{\citenamefont {Veschgini}\ \emph {et~al.}(2011)\citenamefont
  {Veschgini}, \citenamefont {Megias},\ and\ \citenamefont
  {Pirner}}]{Veschgini:2010ws}%
  \BibitemOpen
  \bibfield  {author} {\bibinfo {author} {\bibfnamefont {K.}~\bibnamefont
  {Veschgini}}, \bibinfo {author} {\bibfnamefont {E.}~\bibnamefont {Megias}}, \
  and\ \bibinfo {author} {\bibfnamefont {H.}~\bibnamefont {Pirner}},\ }\Doi
  {10.1016/j.physletb.2011.01.011} {\bibfield  {journal} {\bibinfo  {journal}
  {Phys.Lett.},\ }\textbf {\bibinfo {volume} {B696}},\ \bibinfo {pages} {495}
  (\bibinfo {year} {2011})}\BibitemShut {NoStop}%
\bibitem [{\citenamefont {Boyd}\ \emph {et~al.}(1996)\citenamefont {Boyd},
  \citenamefont {Engels}, \citenamefont {Karsch}, \citenamefont {Laermann},
  \citenamefont {Legeland} \emph {et~al.}}]{Boyd:1996bx}%
  \BibitemOpen
  \bibfield  {author} {\bibinfo {author} {\bibfnamefont {G.}~\bibnamefont
  {Boyd}}, \bibinfo {author} {\bibfnamefont {J.}~\bibnamefont {Engels}},
  \bibinfo {author} {\bibfnamefont {F.}~\bibnamefont {Karsch}}, \bibinfo
  {author} {\bibfnamefont {E.}~\bibnamefont {Laermann}}, \bibinfo {author}
  {\bibfnamefont {C.}~\bibnamefont {Legeland}},  \emph {et~al.},\ }\Doi
  {10.1016/0550-3213(96)00170-8} {\bibfield  {journal} {\bibinfo  {journal}
  {Nucl. Phys.},\ }\textbf {\bibinfo {volume} {B469}},\ \bibinfo {pages} {419}
  (\bibinfo {year} {1996})}\BibitemShut {NoStop}%
\bibitem [{\citenamefont {Narison}(2004)}]{narison2004qcd}%
  \BibitemOpen
  \bibfield  {author} {\bibinfo {author} {\bibfnamefont {S.}~\bibnamefont
  {Narison}},\ }\href@noop {} {\emph {\bibinfo {title} {{QCD as a theory of
  hadrons: from partons to confinement}}}}\ (\bibinfo  {publisher}
  {Cambridge},\ \bibinfo {year} {2004})\BibitemShut {NoStop}%
\bibitem [{\citenamefont {Shifman}(1992)}]{shifman1992vacuum}%
  \BibitemOpen
  \bibfield  {author} {\bibinfo {author} {\bibfnamefont {M.}~\bibnamefont
  {Shifman}},\ }\href@noop {} {\emph {\bibinfo {title} {{Vacuum structure and
  QCD sum rules}}}}\ (\bibinfo  {publisher} {North-Holland},\ \bibinfo {year}
  {1992})\BibitemShut {NoStop}%
\bibitem [{\citenamefont {Lavelle}\ and\ \citenamefont
  {Schaden}(1988)}]{Lavelle:1988eg}%
  \BibitemOpen
  \bibfield  {author} {\bibinfo {author} {\bibfnamefont {M.~J.}\ \bibnamefont
  {Lavelle}}\ and\ \bibinfo {author} {\bibfnamefont {M.}~\bibnamefont
  {Schaden}},\ }\Doi {10.1016/0370-2693(88)90433-9} {\bibfield  {journal}
  {\bibinfo  {journal} {Phys. Lett.},\ }\textbf {\bibinfo {volume} {B208}},\
  \bibinfo {pages} {297} (\bibinfo {year} {1988})}\BibitemShut {NoStop}%
\bibitem [{\citenamefont {Lavelle}\ and\ \citenamefont
  {Schaden}(1990)}]{Lavelle:1990xg}%
  \BibitemOpen
  \bibfield  {author} {\bibinfo {author} {\bibfnamefont {M.}~\bibnamefont
  {Lavelle}}\ and\ \bibinfo {author} {\bibfnamefont {M.}~\bibnamefont
  {Schaden}},\ }\Doi {10.1016/0370-2693(90)90635-J} {\bibfield  {journal}
  {\bibinfo  {journal} {Phys. Lett.},\ }\textbf {\bibinfo {volume} {B246}},\
  \bibinfo {pages} {487} (\bibinfo {year} {1990})}\BibitemShut {NoStop}%
\bibitem [{\citenamefont {Lavelle}\ and\ \citenamefont
  {Schaden}(1989)}]{Lavelle:1988fj}%
  \BibitemOpen
  \bibfield  {author} {\bibinfo {author} {\bibfnamefont {M.}~\bibnamefont
  {Lavelle}}\ and\ \bibinfo {author} {\bibfnamefont {M.}~\bibnamefont
  {Schaden}},\ }\Doi {10.1016/0370-2693(89)90095-6} {\bibfield  {journal}
  {\bibinfo  {journal} {Phys. Lett.},\ }\textbf {\bibinfo {volume} {B217}},\
  \bibinfo {pages} {551} (\bibinfo {year} {1989})}\BibitemShut {NoStop}%
\bibitem [{\citenamefont {Bagan}\ and\ \citenamefont
  {Steele}(1989)}]{Bagan:1989dr}%
  \BibitemOpen
  \bibfield  {author} {\bibinfo {author} {\bibfnamefont {E.}~\bibnamefont
  {Bagan}}\ and\ \bibinfo {author} {\bibfnamefont {T.~G.}\ \bibnamefont
  {Steele}},\ }\Doi {10.1016/0370-2693(89)90303-1} {\bibfield  {journal}
  {\bibinfo  {journal} {Phys. Lett.},\ }\textbf {\bibinfo {volume} {B226}},\
  \bibinfo {pages} {142} (\bibinfo {year} {1989})}\BibitemShut {NoStop}%
\bibitem [{\citenamefont {Lavelle}\ and\ \citenamefont
  {Oleszczuk}(1992)}]{Lavelle:1992yh}%
  \BibitemOpen
  \bibfield  {author} {\bibinfo {author} {\bibfnamefont {M.}~\bibnamefont
  {Lavelle}}\ and\ \bibinfo {author} {\bibfnamefont {M.}~\bibnamefont
  {Oleszczuk}},\ }\Doi {10.1142/S0217732392003049} {\bibfield  {journal}
  {\bibinfo  {journal} {Mod. Phys. Lett.},\ }\textbf {\bibinfo {volume} {A7}},\
  \bibinfo {pages} {3617} (\bibinfo {year} {1992})}\BibitemShut {NoStop}%
\bibitem [{\citenamefont {Schaefer}\ and\ \citenamefont
  {Thoma}(1999)}]{Schaefer:1998wd}%
  \BibitemOpen
  \bibfield  {author} {\bibinfo {author} {\bibfnamefont {A.}~\bibnamefont
  {Schaefer}}\ and\ \bibinfo {author} {\bibfnamefont {M.~H.}\ \bibnamefont
  {Thoma}},\ }\Doi {10.1016/S0370-2693(99)00186-0} {\bibfield  {journal}
  {\bibinfo  {journal} {Phys. Lett.},\ }\textbf {\bibinfo {volume} {B451}},\
  \bibinfo {pages} {195} (\bibinfo {year} {1999})}\BibitemShut {NoStop}%
\bibitem [{\citenamefont {Mustafa}\ \emph
  {et~al.}(2000){\natexlab{a}}\citenamefont {Mustafa}, \citenamefont
  {Schaefer},\ and\ \citenamefont {Thoma}}]{Mustafa:1999jz}%
  \BibitemOpen
  \bibfield  {author} {\bibinfo {author} {\bibfnamefont {M.~G.}\ \bibnamefont
  {Mustafa}}, \bibinfo {author} {\bibfnamefont {A.}~\bibnamefont {Schaefer}}, \
  and\ \bibinfo {author} {\bibfnamefont {M.~H.}\ \bibnamefont {Thoma}},\ }\Doi
  {10.1016/S0370-2693(99)01441-0} {\bibfield  {journal} {\bibinfo  {journal}
  {Phys. Lett.},\ }\textbf {\bibinfo {volume} {B472}},\ \bibinfo {pages} {402}
  (\bibinfo {year} {2000}{\natexlab{a}})}\BibitemShut {NoStop}%
\bibitem [{\citenamefont {Lee}\ \emph {et~al.}(1999)\citenamefont {Lee},
  \citenamefont {Wirstam}, \citenamefont {Zahed},\ and\ \citenamefont
  {Hansson}}]{Lee:1998nz}%
  \BibitemOpen
  \bibfield  {author} {\bibinfo {author} {\bibfnamefont {C.~H.}\ \bibnamefont
  {Lee}}, \bibinfo {author} {\bibfnamefont {J.}~\bibnamefont {Wirstam}},
  \bibinfo {author} {\bibfnamefont {I.}~\bibnamefont {Zahed}}, \ and\ \bibinfo
  {author} {\bibfnamefont {T.~H.}\ \bibnamefont {Hansson}},\ }\Doi
  {10.1016/S0370-2693(99)00061-1} {\bibfield  {journal} {\bibinfo  {journal}
  {Phys. Lett.},\ }\textbf {\bibinfo {volume} {B448}},\ \bibinfo {pages} {168}
  (\bibinfo {year} {1999})}\BibitemShut {NoStop}%
\bibitem [{\citenamefont {Mustafa}\ \emph
  {et~al.}(2000){\natexlab{b}}\citenamefont {Mustafa}, \citenamefont
  {Schafer},\ and\ \citenamefont {Thoma}}]{Mustafa:1999dt}%
  \BibitemOpen
  \bibfield  {author} {\bibinfo {author} {\bibfnamefont {M.~G.}\ \bibnamefont
  {Mustafa}}, \bibinfo {author} {\bibfnamefont {A.}~\bibnamefont {Schafer}}, \
  and\ \bibinfo {author} {\bibfnamefont {M.~H.}\ \bibnamefont {Thoma}},\ }\Doi
  {10.1103/PhysRevC.61.024902} {\bibfield  {journal} {\bibinfo  {journal}
  {Phys. Rev.},\ }\textbf {\bibinfo {volume} {C61}},\ \bibinfo {pages} {024902}
  (\bibinfo {year} {2000}{\natexlab{b}})}\BibitemShut {NoStop}%
\bibitem [{\citenamefont {Schmidt}\ and\ \citenamefont
  {Yang}(1999)}]{Schmidt:1999je}%
  \BibitemOpen
  \bibfield  {author} {\bibinfo {author} {\bibfnamefont {I.}~\bibnamefont
  {Schmidt}}\ and\ \bibinfo {author} {\bibfnamefont {J.-J.}\ \bibnamefont
  {Yang}},\ }\Doi {10.1016/S0370-2693(99)01201-0} {\bibfield  {journal}
  {\bibinfo  {journal} {Phys. Lett.},\ }\textbf {\bibinfo {volume} {B468}},\
  \bibinfo {pages} {138} (\bibinfo {year} {1999})}\BibitemShut {NoStop}%
\bibitem [{\citenamefont {Heinz}\ \emph {et~al.}(1987)\citenamefont {Heinz},
  \citenamefont {Kajantie},\ and\ \citenamefont {Toimela}}]{Heinz:1986kz}%
  \BibitemOpen
  \bibfield  {author} {\bibinfo {author} {\bibfnamefont {U.~W.}\ \bibnamefont
  {Heinz}}, \bibinfo {author} {\bibfnamefont {K.}~\bibnamefont {Kajantie}}, \
  and\ \bibinfo {author} {\bibfnamefont {T.}~\bibnamefont {Toimela}},\ }\Doi
  {10.1016/0003-4916(87)90002-9} {\bibfield  {journal} {\bibinfo  {journal}
  {Ann. Phys.},\ }\textbf {\bibinfo {volume} {176}},\ \bibinfo {pages} {218}
  (\bibinfo {year} {1987})}\BibitemShut {NoStop}%
\bibitem [{\citenamefont {Weldon}(1982)}]{Weldon:1982aq}%
  \BibitemOpen
  \bibfield  {author} {\bibinfo {author} {\bibfnamefont {A.~H.}\ \bibnamefont
  {Weldon}},\ }\Doi {10.1103/PhysRevD.26.1394} {\bibfield  {journal} {\bibinfo
  {journal} {Phys. Rev.},\ }\textbf {\bibinfo {volume} {D26}},\ \bibinfo
  {pages} {1394} (\bibinfo {year} {1982})}\BibitemShut {NoStop}%
\bibitem [{\citenamefont {Spiridonov}(1990)}]{Spiridonov:1990rr}%
  \BibitemOpen
  \bibfield  {author} {\bibinfo {author} {\bibfnamefont {V.}~\bibnamefont
  {Spiridonov}},\ }\Doi {10.1142/S0217732390000731} {\bibfield  {journal}
  {\bibinfo  {journal} {Mod.Phys.Lett.},\ }\textbf {\bibinfo {volume} {A5}},\
  \bibinfo {pages} {653} (\bibinfo {year} {1990})}\BibitemShut {NoStop}%
\bibitem [{\citenamefont {Chakraborty}\ \emph {et~al.}()\citenamefont
  {Chakraborty}, \citenamefont {Mustafa},\ and\ \citenamefont
  {Thoma}}]{cmt:2011b}%
  \BibitemOpen
  \bibfield  {author} {\bibinfo {author} {\bibfnamefont {P.}~\bibnamefont
  {Chakraborty}}, \bibinfo {author} {\bibfnamefont {M.~G.}\ \bibnamefont
  {Mustafa}}, \ and\ \bibinfo {author} {\bibfnamefont {M.~H.}\ \bibnamefont
  {Thoma}},\ }\href@noop {} {}\bibinfo {note} {In preparation}\BibitemShut
  {NoStop}%
\bibitem [{\citenamefont {Reinders}\ \emph {et~al.}(1985)\citenamefont
  {Reinders}, \citenamefont {Rubinstein},\ and\ \citenamefont
  {Yazaki}}]{Reinders:1984sr}%
  \BibitemOpen
  \bibfield  {author} {\bibinfo {author} {\bibfnamefont {L.}~\bibnamefont
  {Reinders}}, \bibinfo {author} {\bibfnamefont {H.}~\bibnamefont
  {Rubinstein}}, \ and\ \bibinfo {author} {\bibfnamefont {S.}~\bibnamefont
  {Yazaki}},\ }\Doi {10.1016/0370-1573(85)90065-1} {\bibfield  {journal}
  {\bibinfo  {journal} {Phys. Rept.},\ }\textbf {\bibinfo {volume} {127}},\
  \bibinfo {pages} {1} (\bibinfo {year} {1985})}\BibitemShut {NoStop}%
\bibitem [{\citenamefont {Wang}(2000)}]{Wang:2000uj}%
  \BibitemOpen
  \bibfield  {author} {\bibinfo {author} {\bibfnamefont {X.-N.}\ \bibnamefont
  {Wang}},\ }\Doi {10.1016/S0370-2693(00)00642-0} {\bibfield  {journal}
  {\bibinfo  {journal} {Phys. Lett.},\ }\textbf {\bibinfo {volume} {B485}},\
  \bibinfo {pages} {157} (\bibinfo {year} {2000})}\BibitemShut {NoStop}%
\bibitem [{\citenamefont {Matsui}\ and\ \citenamefont
  {Satz}(1986)}]{Matsui:1986dk}%
  \BibitemOpen
  \bibfield  {author} {\bibinfo {author} {\bibfnamefont {T.}~\bibnamefont
  {Matsui}}\ and\ \bibinfo {author} {\bibfnamefont {H.}~\bibnamefont {Satz}},\
  }\Doi {10.1016/0370-2693(86)91404-8} {\bibfield  {journal} {\bibinfo
  {journal} {Phys. Lett.},\ }\textbf {\bibinfo {volume} {B178}},\ \bibinfo
  {pages} {416} (\bibinfo {year} {1986})}\BibitemShut {NoStop}%
\end{thebibliography}%

\end{document}